\documentclass[aps,10pt,nofootinbib,prd,singlecolumn,showpacs,superscriptaddress,groupedaddress]{revtex4-2}

\usepackage[american]{babel}
\usepackage{amsfonts}
\usepackage{amsmath}
\usepackage{amssymb}
\usepackage{slashed}
\usepackage{xcolor}
\usepackage{bm}
\usepackage{float}
\usepackage[utf8]{inputenc}
\usepackage[macrosonly]{chet}
\usepackage{nicefrac}
\usepackage{hyperref}
\usepackage{url}
\usepackage[normalem]{ulem}
\usepackage{lmodern}
\usepackage{tikz}
\usepackage{mathtools}
\usetikzlibrary{decorations.pathmorphing}

\tikzset{graviton/.style={decorate, decoration={snake, segment length=2mm, amplitude=0.6mm}}}

\newcommand{\dd}{{\rm d}}
\newcommand{\tr}{{\rm tr}}

\newcommand{\Tad}{\textbf{\text{Tad}}}
\newcommand{\UVB}{\textbf{\text{UVB}}}
\newcommand{\IRB}{\textbf{\text{IRB}}}
\newcommand{\UVBgh}{\textbf{\text{UVB}}_\text{gh}}
\newcommand{\UVBY}{\textbf{\text{UVB}}_Y}

\begin{document}

	\title{Gauge dependence of momentum running in higher-derivative gravity}
	
	\author{Diego Buccio}
	\email{diego.buccio@ru.nl}
	\affiliation{Institute for Mathematics, Astrophysics and Particle Physics (IMAPP), Radboud University, Heyendaalseweg 135, 6525 AJ Nijmegen, The Netherlands}
	
	\author{Gustavo P. de Brito}
	\email{gp.brito@unesp.br}
	\affiliation{Departamento de F\'isica, Universidade Estadual Paulista (Unesp), Campus Guaratinguet\'a, Av.~Dr.~Ariberto Pereira da Cunha, 333, Guaratinguet\'a, SP, Brazil.}
    
        \author{Luca Parente}
	\email{luca.parente@phd.unipi.it}
	\affiliation{Universit\`a di Pisa and INFN - Sezione di Pisa, Largo Bruno Pontecorvo 3, 56127 Pisa, Italy}

	\begin{abstract}
	Recent works \cite{Buccio:2024hys,Buccio:2024omv} have argued that improved one-loop beta-functions capturing the physical momentum dependence of one-loop corrected higher-derivative gravity theories are the most suitable to describe their high-energy behaviour. 
    This work critically tests the validity of this claim.
        We compute the explicit gauge dependence of the one-loop momentum running of curvature-squared operators in quadratic gravity and conformal gravity using the background field method. We find them to be gauge dependent, and we discuss the implications of this result for the theory and its physical predictivity.
	\end{abstract}

	\maketitle

    \section{Introduction}\label{sect:intro}
	
	In two previous works by some of the authors \cite{Buccio:2024hys,Buccio:2024omv}, it was claimed that the most appropriate one-loop beta functions to describe the high-energy behaviour of higher-derivative theories, and in particular of quadratic gravity and conformal gravity, are not those identified by UV divergences, but those defined by the logarithmic momentum-dependent form factors in the effective action. While these two definitions give identical results in standard theories with $2$-derivative kinetic terms, the same is not true in higher derivative ones \cite{Buccio:2023lzo}.
    See also \cite{Kawai:2025wkp} for a discussion on the relation between the different definitions of running and possible redefinitions of the scale $\mu$ introduced in dimensional regularization.
	
	This difference is the result of on-shell infrared divergences that emerge in the large momentum limit, which cannot be treated with the usual machinery introduced for soft and collinear IR divergences \cite{Weinberg:1965nx}.
	In fact, after a partial fraction decomposition of the propagator, one can show that the same infrared divergences can be seen as sub-subleading logarithmic terms of the UV divergences in a theory with higher-derivative ghosts \cite{Buccio:2025his,Salvio:2025cmi}. 
	
	Let us take a $4$-derivative theory with some interaction vertex $\mathcal{O}$. Differently from a $2$-derivative theory, the Feynman integral associated with the bubble diagram including two copies of the vertex is potentially IR divergent, if the vertex $\mathcal{O}$ itself does not contain powers of the integrated momentum $q$, i.e.,
	\begin{equation}
		\int d^4q\, \frac{\mathcal{O}^2}{(q^4+q^2m^2)\left[(q+p)^4+m^2(q+p)^2\right]}\xrightarrow{p^2\gg m^2} \frac{2\pi^2\mathcal{O}^2}{p^4}\log\left(\frac{p^2}{m^2}\right)\ .
	\end{equation}
	Alternatively, the  propagator can be decomposed via a partial fraction into two 2-derivative propagators with opposite signs in the following way
	\begin{equation}
		\frac{1}{p^4+(m_1^2+m_2^2)p^2+m_1^2m_2^2}=\frac{1}{(p^2+m_1^2)(p^2+m_2^2)}=\frac{1}{m_1^2-m_2^2}\left[\frac{1}{p^2+m_2^2}-\frac{1}{p^2+m_1^2}\right]\, . \label{HDprop}
	\end{equation} 
	Let us now call $I_2(m_a,m_b,p)$ the integral associated to the bubble diagram with standard 2-derivative propagators and different masses, regulated by some UV cutoff $\Lambda$
	\begin{equation}
		I_2(m_a,m_b,p)=\int^\Lambda d^4q\, \frac{\mathcal{O}^2}{(q^2+m_a^2)\left[(q+p)^2+m_b^2\right]}\, .
	\end{equation}
	Then, the 4-derivative bubble will be
	\begin{equation}
		I_4(m_1,m_2,p)=(m_1^2-m_2^2)^{-2}\left[I_2(m_1,m_1,p)-I_2(m_1,m_2,p)-I_2(m_2,m_1,p)+I_2(m_2,m_2,p)\right]\ .\label{sum2D}
	\end{equation}
	In the relevant limits of large and small momentum $p$, the integral $I_2$ is
	\begin{equation}
		I_2(m_1,m_2,p)\sim
		\begin{cases}
			2\mathcal{O}^2\left[\log\left(\frac{p^2}{\Lambda^2}\right)+\sum_i \frac{m_i^2}{p^2}\log\left(\frac{m_i^2}{p^2}\right)+\frac{m_1^2m_2^2}{p^4}\sum_i\log\left(\frac{m_i^2}{p^2}\right)+O\left(\frac{p^4}{m_i^4}\right)\right]
			\ &\text{if}\ p^2\gg m_i^2\, ;
			\\[+1ex]
			\frac{1}{m_1^2-m_2^2}\left[ m_1^2\log\left(\frac{m_1^2}{\Lambda^2}\right)-m_2^2\log\left(\frac{m_2^2}{\Lambda^2}\right)\right]+O\left(\frac{p}{m_i}\right)  &\text{if}\  p^2\ll m_i^2\, ,
		\end{cases}
	\end{equation}
	which is UV divergent, as expected, but also IR finite in the limit of small $p$. It is important to notice that we recover the well-known $\log\left(\frac{p^2}{\Lambda^2}\right)$
	at leading order of the high-energy limit, but also the subleading correction $\log\left(\frac{p^2}{m_i^2}\right)$ starts to appear.
	When we move to $I_4$, the leading and subleading terms of the high energy limit cancel against each other, and we remain with
	\begin{align}
		I_4(m_1,m_2,p)\sim
		\begin{cases}
			\mathcal{O}^2\frac{1}{m_1^2-m_2^2}\sum_i (-1)^i\frac{m_i^2}{p^4}\log\left(\frac{m_i^2}{p^2}\right)+O\left(\frac{p^4}{m_i^4}\right)\ & \text{if}\ p^2\gg m_i^2\, ;\\[+2ex]
			2-2\frac{m_1^2+m_2^2}{m_1^2-m_2^2}\log\left(\frac{m_1}{m_2}\right) +O\left(\frac{p}{m_i}\right) \qquad &\text{if}\  p^2\ll m_i^2\, .
		\end{cases}
	\end{align}
	This means that the infrared large logarithms $\log\left(\frac{m_i^2}{p^2}\right)$ in higher-derivative theories do not come from infrared divergences regulated by the masses in the 2-derivative terms of the sum (\ref{sum2D}), but rather from their sub-subleading terms in the high-energy limit, hence they cannot be treated as soft or collinear ones. 
    
	In the present work, we compute the gauge dependence of the momentum induced running using the background field formalism \cite{DeWitt:1967ub,Weinberg:1996kr}, both for the beta functions associated with local covariant operators and for those related to the nonlocal partners of the Einstein-Hilbert term \cite{Donoghue:2022chi}.
	The gauge dependence of momentum-induced beta functions of quadratic gravity has been computed independently in a recent work \cite{Salvio:2025cmi} with a partially alternative approach.

	The rest of the paper is structured as follows.
	In Section~\ref{sect:frame}, we briefly present the theories of quadratic gravity and conformal gravity and introduce the background field method, together with our choice for the gauge parameters. 
	In Section~\ref{sect:gauge}, we present the gauge-dependent beta functions for the operators quadratic in curvature with at most one inverse power of the $\square$ operator.
	In Section~\ref{sect:comment}, we analyse this gauge dependence and its source within the background field method.
	In Section~\ref{sect:conclusions}, we summarize the obtained results and discuss future perspectives in the study of quadratic gravity. Throughout this paper, we work with Euclidean signature. In this way, we avoid possible issues with Wick rotation in the presence of higher-derivative poles \cite{Aglietti:2016pwz, Buoninfante:2025klm}.
	
    \section{The framework}\label{sect:frame}
	
	In this section, we present the main aspects of the framework considered in this paper. Our goal is to compute the one-loop leading-log corrections to the effective action of quadratic gravity and its symmetry-enhanced counterpart called conformal gravity.
	
	We start with quadratic gravity \cite{Stelle:1976gc}. The classical action of quadratic gravity in the Riemann basis is
	\begin{equation}\label{eq:action-qg}
		\begin{split}
			S_{\rm qg}[g_{\mu\nu}]= & \int \dd^4 x \sqrt{g} \,\Bigl[ -Z_N R+
			\alpha R^2+\beta R^{\mu\nu}R_{\mu\nu}+\gamma R^{\mu\nu\rho\sigma}R_{\mu\nu\rho\sigma} 
			\Bigr] 
			\,,
		\end{split}
	\end{equation}
        $\alpha$, $\beta$ and $\gamma$ are dimensionless couplings, and $Z_N$ is related to the Newton coupling $G_N$ through $Z_N = (16\pi G_N)^{-1}$.
        Note that we are setting the cosmological constant to zero to be consistent with the choice of an asymptotically flat background. For our purposes, it is convenient to rewrite the action of quadratic gravity in the Weyl basis, namely
	\begin{equation}\label{eq:action-qgW}
		\begin{split}
			S_{\rm qg}[g_{\mu\nu}]= & \int \dd^4 x \sqrt{g} \, \Bigl[\,-Z_N R+
			\frac{1}{\xi} R^2+\frac{1}{2\lambda}C^{\mu\nu\rho\sigma}C_{\mu\nu\rho\sigma} 
			\,\Bigr]
			\,,
		\end{split}
	\end{equation}
        where the new dimensionless couplings $\xi$ and $\lambda$ are related to the original ones according to
        \begin{equation}
            \lambda = \frac{1}{\beta+4\gamma} 
            \quad \text{and} \quad
            \xi = \frac{3}{3 \alpha + \beta + \gamma} \,.
        \end{equation}
	 In the latter form, we also neglected the topological Gauss-Bonnet term, which is irrelevant for our discussion.
    
    Quadratic gravity is perturbatively renormalizable \cite{Stelle:1976gc}, and that makes it an interesting candidate as a UV completion of general relativity. However, since the gravitonal perturbations have 4-derivative kinetic terms, the theory suffers from Ostrogradsky unbounded energy at the classical level \cite{Ostrogradsky:1850fid}. Surprisingly, it has been recently shown that an unbounded energy does not always lead to classical instability in higher derivative theories \cite{Deffayet:2023wdg,Held:2025fii}.
    This feature leads to an unbounded energy spectrum in the canonically quantized version of the theory, and to problems with unitarity, negative probabilities, or recovering classical gravitation when non-canonical quantizations are employed \cite{Woodard:2015zca}.
    In the past years, some alternative approaches toward a consistent quantization procedure have been suggested (see, for example, \cite{Anselmi:2018ibi,Mannheim:2020ryw,Donoghue:2021eto, Salvio:2015gsi,Holdom:2015kbf}). In most cases, the price to be paid is a loss of microcausality at transplanckian scales.
    
	
    In the second case of study, we consider conformal gravity, which is a kind of symmetry-enhanced version of quadratic gravity where, besides diffeomorphism invariance, the classical action is also symmetric under Weyl transformation $g_{\mu\nu}(x) \mapsto \Omega^2(x) g_{\mu\nu}$, with $\Omega^2(x)$ being a local scalar function. The action of conformal gravity is given by
	\begin{equation}\label{eq:action-cg}
		\begin{split}
			S_{\rm cg}[g_{\mu\nu}]= & \int \dd^4 x \sqrt{g} \, \Bigl[\, \frac{1}{2\lambda}C^{\mu\nu\rho\sigma}C_{\mu\nu\rho\sigma} \, 
			\Bigr]
			\,.
		\end{split}
	\end{equation}
    
	In the following, we employ the background field method to compute the so-called leading-log contributions to the one-loop effective action. This is based on the decomposition of the full metric in terms of a background metric $\bar{g}_{\mu\nu}$ and a fluctuation field $h_{\mu\nu}$. We then integrate over $h_{\mu\nu}$ to obtain the effective dynamics.
	Here, we consider the usual linear decomposition defined as
	\begin{equation}\label{eq:linear_split}
		\begin{split}
			g_{\mu\nu}=\bar{g}_{\mu\nu}+h_{\mu\nu}
			\, .
		\end{split}
	\end{equation}
	To evaluate one-loop contributions to the effective action $\Gamma$, we need to expand the action up to second order in the fluctuation field $h_{\mu\nu}$. The important contribution to our calculation is captured by
	\begin{equation}
		S^{(2)}[\bar{g}_{\mu\nu},h_{\mu\nu}]= \frac{1}{2} 
		\int d^4x \sqrt{\bar{g}} \,  h_{\mu\nu}\frac{\delta^2 S}{\delta g_{\mu\nu}\delta g_{\rho\sigma}}\bigg|_{g=\bar{g}} h_{\rho\sigma}\ .
	\end{equation}
	This step has been done in previous works (see, \textit{e.g.}, \cite{Percacci:2017fkn}), thus we will not report the explicit expressions here.
	
	Due to diffeomorphism symmetry, we need to choose a gauge-fixing condition to remove redundant field configurations from the path integral of $h_{\mu\nu}$. 
	Here, we will consider a fourth-derivative gauge-fixing term which makes the scaling of the graviton propagator homogeneous in the deep UV. In practice, we add the following term to the action of quadratic gravity (see, for instance, \cite{Ohta:2016jvw})
	\begin{equation}\label{eq:GF+FP}
		S_{\rm GF+FP+LN}[h,\bar{g}]=-\int \dd^4 x \sqrt{\bar g} \, \bigg( \frac{1}{2 g_1} F_\mu Y^{\mu\nu} F_\nu
		+  \bar c_\mu \,Y^{\mu}{}_\rho\big[ \, \bar g^{\rho\nu} \bar\Box +(2 g_2+1)\bar\nabla^\rho \bar\nabla^\nu +\bar R^{\rho\nu} \,\big] c_\nu+\frac{1}{2}b_\mu Y^{\mu\nu} b_\nu
		\bigg)\,,
	\end{equation}
	where,
	\begin{equation}\label{eq:gauge}
		\begin{split}
			F_\mu=\bar\nabla^\lambda h_{\lambda\mu}+g_2\bar\nabla_\mu h^{\nu}{}_\nu \qquad
			\text{and}\qquad
			Y_{\mu\nu}=\bar g_{\mu\nu}\bar\Box+g_3\bar\nabla_\mu\bar\nabla_\nu-g_4\bar\nabla_\nu\bar\nabla_\mu
			\,.
		\end{split}
	\end{equation}
	Geometrical objects with an over-bar (such as $\bar\nabla$, $R_{\mu\nu}$, and so on) are computed with respect to the background metric $\bar{g}_{\mu\nu}$.
	We use $c_\mu$ and $\bar{c}_\mu$ to denote the pair of Faddeev-Popov (FP) ghosts, while $b_\mu$ is the Lautrup-Nakanishi (LN) field. The latter contributes non-trivially to the effective action due to the derivative operators appearing on $Y^{\mu\nu}$.
	The parameters $g_1$, $g_2$, $g_3$, and $g_4$ are gauge-fixing parameters, which in principle can be arbitrarily chosen.
	Calculations of quantum corrections to the effective action in quadratic gravity are commonly done with the so-called minimal gauge. This gauge choice is characterized by
	\begin{equation}\label{eq:MinGFparameters}
		\begin{split}
			g_1=\lambda\,, \qquad g_2=-\frac14 +\frac{9\lambda}{4(\xi-3\lambda)}\,, \qquad g_3=\frac23-\frac{2\lambda}{\xi} \,, \qquad g_4=1
			\,.
		\end{split}
	\end{equation}
	The minimal gauge is chosen such that the $4$-derivative gauge-fixed Hessian takes the simple form $\square^2$, see, e.g., Ref.~\cite{Ohta:2013uca}. The minimal gauge was adopted in \cite{Buccio:2024hys} as part of the computation of running couplings in terms of leading-log contributions. In the present paper, we keep the gauge parameters $g_1$, $g_2$, $g_3$, and $g_4$ completely arbitrary in order to study the gauge-dependence of the leading-log contributions to the effective action.

	In the case of conformal gravity, we also have Weyl invariance, so the redundancies due to this symmetry can be explicitly fixed by projecting the quantum fluctuation onto its traceless part. To do this, we take $g_2=-1/4$ in the gauge fixing function $F_\mu$ for conformal gravity.

    Thanks to the background-field method, the effective action as a functional of the background metric $\bar{g}_{\mu\nu}$ (with $h_{\mu\nu}=0$) is a manifestly covariant functional with structure
    \cite{Barvinsky:1990up,Codello:2012kq}
        \begin{equation}\label{eq:action-form-factors}
		\Gamma[\bar{g}] = \int \dd^4 x \,\sqrt{\bar{g}} \, \Big[
            - Z_N \bar{R} +
		\bar{C}_{\mu\nu\alpha\beta} f_{\lambda}(\bar{\Box};\mu^2,Z_N) \bar{C}^{\mu\nu\alpha\beta}
		+ \bar{R} f_{\xi}(\bar{\Box};\mu^2,Z_N) \bar{R}\,
		\Big] +{\cal O}({\cal R}^3)
		\,,
	\end{equation}
	where $\mathcal{R}$ represents any generic curvature tensor. The momentum running and the associated beta functions are defined by the coefficients of the parts proportional to $\log\bar{\square}$ in the one-loop corrections to the functions $f_\lambda$ and $f_\xi$ in the high-energy limit.
    
	The effective action is obtained by integrating out $h$, $\bar{c}$, $c$, and $b$ in the path integral, and the one-loop correction is given by \cite{Avramidi:1985ki,Fradkin:1981iu,Ohta:2013uca}
	\begin{equation}\label{eq:effective-action-CG}
		\begin{split}
			\Delta\Gamma_\text{one-loop} =  \frac{1}{2} \tr \log H-\tr \log \Delta_{\rm gh}-\frac12\tr \log Y
			\,,
		\end{split}
	\end{equation}
	where we introduced the modified ghost differential operator
	\begin{equation}\label{eq:FP}
		\begin{split}
			(\Delta_{\rm gh})^{\mu}{}_\nu = \delta^\mu_\nu \, \bar\Box+(2g_2+1)\,\bar\nabla^\mu\bar\nabla_\nu+\bar R^\mu{}_\nu
			\, ,
		\end{split}
	\end{equation}
	and the gauge-fixed Hessian
	\begin{equation}
		H^{\mu\nu\alpha\beta} = 
		\frac{\delta^2 (S[\bar{g}+h] + S_{\rm GF}[h,\bar{g}])}{\delta h_{\mu\nu}\delta h_{\alpha\beta}}\Bigg|_{h=0}\, .
	\end{equation}
	The problem of calculating the UV divergences of these functional traces is an old one, but it can be accomplished using the general results of \cite{Barvinsky:1985an}.

	The computation of the IR contributions is considerably more challenging.
	To compute the IR divergent parts, which, as discussed above, occur only in the functional traces of higher derivative operators, we adopt the strategy that has already been described in \cite{Buccio:2024hys,Buccio:2024omv} and mutuated from \cite{tHooft:1974toh,Julve:1978xn}. We also expand the background metric around flat spacetime as $\bar{g}_{\mu\nu}=\delta_{\mu\nu}+f_{\mu\nu}$, so that we can compute the background's 2-point function 
	\begin{equation}
		\langle f^{\mu\nu}f^{\alpha\beta} \rangle = \frac{\delta^{2} \Gamma[\bar{g} =\delta +f]}{\delta f_{\mu\nu} \delta f_{\alpha\beta}}\Bigg|_{f=0} \,.
	\end{equation}
	Since the effective action is covariant in the background metric, this 2-point function should allow us to extract the running of operators quadratic in background curvatures in the spirit of the background field method.
	After this expansion, we can introduce a Fourier transform on the flat background and use the standard Feynman diagrams machinery to find both UV and IR divergences, the latter being our main interest.
	
	With the structure of the gauge sector described above, the contributions of the FP ghosts and the LN field are functional traces of second-order operators, so they have only UV divergences, which can be computed following the procedures described, for example, in \cite{Barvinsky:1985an, Avramidi:1985ki}. We will denote them as $\UVBgh$ and $\UVBY$, respectively.

	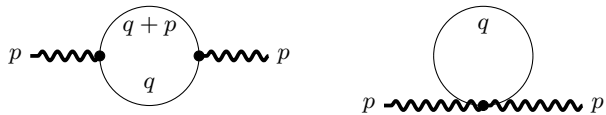
\begin{figure}[ht]
		\begin{center}
			\begin{tikzpicture}
				[>=stealth,scale=0.66,baseline=-0.1cm]
				\draw (-1,0) arc (180:0:1cm);
				\draw (-1,0) arc (180:360:1cm);
				\draw [line width=0.5mm, graviton ] (-1,0) node[anchor=west] {} -- (-2.4,0);
				\draw [line width=0.5mm, graviton ] (2.4,0) -- (1,0) node[anchor=east] {};
				\draw[fill] (1,0) circle [radius=0.1];
				\draw[fill] (-1,0) circle [radius=1mm];
				\node [] at (-2.7,0) {$p$};
				\node [] at (2.7,0) {$p$};
				\node [] at (0,0.6) {$q+p$};
				\node [] at (0,-0.6) {$q$};
			\end{tikzpicture}
			\qquad
			\begin{tikzpicture}
				[>=stealth,scale=0.66,baseline=-0.1cm]
				\draw (-1,0) arc (180:0:1cm);
				\draw (-1,0) arc (180:360:1cm);
				\draw [line width=0.5mm, graviton ](2,-1) node[anchor=west] {} -- (0,-1);
				\draw [line width=0.5mm, graviton ](0,-1) -- (-2,-1) node[anchor=east] {};
				\draw[fill] (0,-1) circle [radius=1mm];
				\node [] at (-2.3,-1) {$p$};
				\node [] at (2.3,-1) {$p$};
				\node [] at (0,0.6) {$q$};
			\end{tikzpicture}
			\caption{Diagrams contributing to the 2-point function:
				bubbles (left) and tadpoles (right).
				The thin line can be the $h$ propagator or one of the ghosts, the thick line is the $f$ propagator, with momentum $p$.
			}
			\label{fig:feynman}
		\end{center}
	\end{figure}
	
	As for the fluctuations $h_{\mu\nu}$, the story is more complicated.
	The $1$-loop contribution to the 2-point function can be separated into tadpole diagrams and bubble diagrams depicted in Figure~\ref{fig:feynman}.
	In higher derivative (HD) theories, these diagrams can be both UV and IR divergent.
	The sum of UV divergences is used to define the ``traditional'' beta functions, which we have referred to as $\mu$-running in previous works \cite{Buccio:2023lzo,Buccio:2024hys,Buccio:2024omv}, where $\mu$ is, for example, the scale introduced by minimal subtraction. They are identified by the $\log\mu$ terms in dimensional regularization in theories where the IR part of loop integrals is regularized by a mass (e.g., the EH term in quadratic gravity). 
	In this scenario, the IR log-divergences are instead identified by $\log m$ factors.
	In the background 2-point function, there is only one external momentum $p^\mu$ due to overall momentum conservation, so the only possible logarithm containing momenta in the high-energy limit ($p^2\gg m^2$) is $\log p^2$. We extract the momentum running, or $p$-running, from the $\log p^2$ terms.
	
	The tadpole loop integral on the right of Figure~\ref{fig:feynman} is independent of $p$, hence it can only produce $\log \frac\mu m$. This means that the UV and IR divergent parts of the tadpole are equal with opposite signs. This is confirmed by the well-known fact that in a scaleless theory, i.e., one for which $m=0$, the tadpole diagram in dimensional regularization is zero. If the integral is dimensionally regulated both in the UV and IR, then $m$ is replaced with $\mu$ and  $\log \frac\mu\mu=0$.
	
	The bubble integral depends instead on $p$ and can produce $\log p$ terms, so there is no such trivial relation between its UV and IR divergent parts.
	In the following, we will divide the one-loop contribution to the effective action into three terms: the UV bubble, the IR bubble, and the tadpole. We will call  $\UVB$ the coefficient in front of $\log\mu$ in the UV bubble,  $\Tad$ the same coefficient in the tadpole, and $\IRB$ the coefficient of $\log m$ in the IR part of the bubble diagram. Given these definitions, in the graviton part of the functional trace, the contribution to the $\mu$-running is given by the sum $\UVB+\Tad$, while the quantity $-(\UVB+\IRB)$, which precisely corresponds to the coefficient of $\log|p|$, contributes to the $p$-running. Finally, the sum of IR divergences is the difference $\IRB-\Tad$. It is quite easy to check that, in this way, the arguments of all logarithms can be made dimensionless. Furthermore, each of these three linear combinations must be covariant, while the same is not true for $\UVB$, $\IRB$, and $\Tad$ taken independently.

    \section{Gauge dependence}\label{sect:gauge}
    In this section, we display the dependence of the different beta functions on the gauge parameters. In particular, we will show that the beta function encoding the running with respect to the momentum $p$, as defined in \cite{Buccio:2024hys,Buccio:2024omv}, turns out to be explicitly gauge dependent.
    
    From now on, all the geometrical objects correspond to quantities evaluated with respect to the background metric $\bar{g}_{\mu\nu}$. For simplicity of notation, from this point on, we will no longer use over-bars to denote geometrical objects evaluated with respect to the background metric.
	
    \subsection{Quadratic Gravity}
	In quadratic gravity, the corrections to the $C^2 (=C_{\mu\nu\alpha\beta} C^{\mu\nu\alpha\beta})$ and $R^2$ terms in the $1$-loop effective action multiplied by $\log\mu$ are
	\begin{equation}
		\Big[\UVB + \Tad  + \UVBgh + \UVBY\Big]_{Z_N=0} 
		=-\frac{5 \left(72 \lambda ^2-36 \lambda  \xi +\xi ^2\right)}{576 \pi ^2 \xi ^2} R^2 -\frac{133
			}{320 \pi ^2}\, C^2\ .\label{muQG}
	\end{equation}
	This expression is gauge-independent and in agreement with the literature \cite{Avramidi:1985ki,deBerredo-Peixoto:2004cjk,Ohta:2013uca}. 
    The beta functions encoding the $\mu$-running of $\lambda$ (and $\xi$) are defined as $2\lambda^2$ (and $\xi^2$) times the coefficient of $C^2$ (and $R^2$) in the last expression, namely
    \begin{subequations}
        \begin{align}
            &\mu \frac{\partial\lambda(\mu)}{\partial \mu} \coloneq \beta_{\lambda}^{(\mu)} = - \frac{1}{(4 \pi)^2} \frac{133 \lambda^2}{10} \,, \\
            &\mu \frac{\partial\xi(\mu)}{\partial \mu} \coloneq \beta_{\xi}^{(\mu)} = - \frac{1}{(4 \pi)^2} \frac{5\left(72 \lambda ^2-36 \lambda  \xi +\xi ^2\right)}{36} \,.
        \end{align}
    \end{subequations}
	
    At leading order in the $p^2\gg Z_N$ limit, the quantum corrections to the covariant structures in (\ref{eq:action-form-factors}) that generate terms multiplied by $\log |p|$ in the momentum-space 2-point function are instead
	\begin{equation}
		\begin{aligned}
		-\Big[\UVB+\IRB+\UVBgh+\UVBY\Big]_{Z_N=0} &=  
            \frac{\mathcal{C}_{R^2}}{20736 \pi ^2 (g_2+1)^2 \,\xi^2} \, R \log(-\square) R  \\
            &-\frac{\mathcal{C}_{C^2}}{1244160 \pi ^2 (g_2+1)^2 (g_3-g_4+1) \,\lambda^2}\, C_{\mu\nu\alpha\beta} \log(-\square) C^{\mu\nu\alpha\beta} \,, \label{p-running}
		\end{aligned}
	\end{equation}
	with gauge-dependent coefficients $\mathcal{C}_{R^2}$ and $\mathcal{C}_{C^2}$ given by
	\begin{equation}
		\begin{aligned}
			 \mathcal{C}_{R^2} &=
			 	\frac{18 g_1 (-120 \lambda  (8 g_2 (g_2+2) (g_3-g_4+1)+8
			 	g_3-8 g_4+9)}{g_3-g_4+1} \\
		 		&+ \frac{44 \xi  (g_2 (g_2+2) (g_3-g_4+1)+g_3-g_4)+45
		 		\xi )}{g_3-g_4+1} \\
			 	&-22320 (g_2+1)^2 \lambda ^2-120 (8 g_2 (4 g_2+5)+17) \lambda  \xi +(2 g_2 (89 g_2+112)+55) \xi ^2 \,,
		\end{aligned} 
	\end{equation}
	and
	\begin{equation}
		\begin{aligned}
			\mathcal{C}_{C^2} &=\frac{}{} 
			90 g_1 (\xi  (8 g_2 (g_2+2) (g_3-g_4+1)+8 g_3-8 g_4+9)\\
			&\frac{}{} -12 \lambda  (26 g_2 (g_2+2) (g_3-g_4+1) +26 g_3-26 g_4+27))\\
			&\frac{}{} +(g_3-g_4+1) \left(-305352 (g_2+1)^2 \lambda ^2+60 (2 (7-16 g_2)
			g_2+37) \lambda  \xi +5 (4 g_2 \xi +\xi )^2\right)  \,.
		\end{aligned}
	\end{equation}
    Here, the beta functions encoding the $p$-running of $\lambda$ (and $\xi$) are defined as $-2\lambda^2$ (and $-\xi^2$) times the coefficient of $C^2$ (and $R^2$) in expression \eqref{p-running}, namely
        \begin{subequations}
            \begin{align}
                &p \frac{\partial \lambda(p)}{\partial p} \coloneq \beta^{(p)}_\lambda =
                \frac{\mathcal{C}_{C^2}}{311040 \pi ^2 (g_2+1)^2 (g_3-g_4+1)} \,,\\
                &p \frac{\partial \xi(p)}{\partial p} \coloneq \beta^{(p)}_\xi =
                -\frac{\mathcal{C}_{R^2}}{10368 \pi ^2 (g_2+1)^2} \,.
            \end{align}
        \end{subequations}
    We notice that, with the gauge choice
    \begin{equation}
        g_2\to \frac{-18 g_1-30 \lambda +\xi }{2 (9 g_1+15 \lambda +\xi
   )},\qquad g_3\to \frac{-2 g_1+g_4 \xi -\xi }{\xi }\ ,
    \end{equation}
    the $p$-running coincides with the $\mu$-running.
    
    In the minimal gauge, see Eq. \eqref{eq:MinGFparameters}, the coefficients of terms proportional $\log|p|$ reduce to
	\begin{equation}\label{minimalph}
		\begin{aligned}
		    -\Big[\UVB+\IRB+\UVBgh+\UVBY\Big]_{Z_N=0} ^\text{min.~gauge} &= 
		\frac{5 \left(\xi^2 -36 \lambda  \xi - 2520 \lambda ^2\right)}{5760 \pi ^2 \xi ^2} \,R \log(-\square) R \\
		&+\frac{(1617 \lambda-20 \xi )}{5760 \pi ^2   \lambda} \,C_{\mu\nu\alpha\beta} \log(-\square) C^{\mu\nu\alpha\beta} \,,
		\end{aligned}
	\end{equation}
    and, therefore, the corresponding $p$-running beta functions reduce to
        \begin{subequations}
            \begin{align}
            &\beta^{(p)}_\lambda \big|_\text{min. gauge} =
            - \frac{1}{(4\pi)^2}\frac{(1617 \lambda-20 \xi ) \lambda}{90} 
             \\
            &\beta^{(p)}_\xi \big|_\text{min. gauge} = - \frac{1}{(4\pi)^2}
            \frac{\xi^2 -36 \lambda  \xi - 2520 \lambda ^2}{36}
        \end{align}
        \end{subequations}
    which is in accordance with the results of Ref.~\cite{Buccio:2024hys}.

    Once stated the gauge dependence of the momentum running of the local operators in quadratic gravity, we would like to see whether it is possible to define a gauge invariant running for the masses of the spin-2 ghost and the spin-0 mode, which are, respectively, $m_2^2=\lambda Z_N$ and $m_0^2=-\xi Z_N/6$ in the classical theory.
    It is well known that the beta function encoding the $\mu$-running of the Einstein-Hilbert term is gauge dependent \cite{Avramidi:1986mj,Shapiro:1994vb}, hence a gauge-independent beta function for the mass parameters is missing. We want to understand whether such a beta function can be defined within the momentum running approach.
	
    The subleading terms in the high-energy limit $p^2\gg Z_N$ of the background 2-point function correspond to the quantum corrections to the Einstein-Hilbert term and its nonlocal partners, 
    \begin{equation}
        R\,\bigg(\frac{1}{-\square}\bigg) R 
        \qquad \text{and}  \qquad 
        C^{\mu\nu\rho\sigma}\bigg(\frac{1}{-\square}\bigg) C_{\mu\nu\rho\sigma} \,.
    \end{equation}

    At one-loop, not only the EH term, but also the nonlocal partners, contribute to the masses of the ghost and of the scalar mode.
    The gauge-fixing functional $F$ and the operator $Y$ that we have chosen are independent of $Z_N$, so the functional traces from path integrals over ghost modes cannot contribute to the running of these operators via $\UVBgh$ and $\UVBY$.
	
    The UV divergences can only contribute via local terms, so they are proportional to $R$. In fact, the UV divergent part proportional to $Z_N$ is
	\begin{equation}
		\big[\UVB+\Tad\big]_{Z_N}=-\frac{Z_N}{4608 \pi ^2}\,\left[ 18 g_1 \left(\frac{1}{(g_2+1)^2 (g_3-g_4+1)}+12\right)+\frac{(4 g_2 (5
			g_2+1)-7) \xi }{(g_2+1)^2}+\frac{1440 \lambda ^2}{\xi }\right] R .
	\end{equation}
	In the minimal gauge, we recover the well-known result, see, e.g., Ref.~\cite{Avramidi:1985ki}
	\begin{equation}
		\big[\UVB+\Tad\big]^\text{min.~gauge}_{Z_N} = -\frac{Z_N (30 \lambda -\xi ) (4 \lambda +\xi )}{384 \pi ^2 \xi }\, R  \, .
	\end{equation}
	No momentum-dependent form factors can be associated with the EH term, in which case the $p$-running can only correspond to a running of the nonlocal partners, that is,
	\begin{equation}
		\begin{aligned}
			\big[\UVB-\IRB\big]_{Z_N} 
			&= \frac{ Z_N \, \mathcal{C}_{R^2}^{Z_N}}{1327104 \pi ^2 (g_2+1)^4 \xi ^2 (g_3-g_4+1)} \,R \left( \frac{\log(-\square)}{-\Box}  \right) R \\
			& + \frac{ Z_N \,\mathcal{C}_{C^2}^{Z_N}}{47775744 \pi ^2 (g_2+1)^4 \lambda ^2 (g_3-g_4+1)} C_{\mu\nu\alpha\beta}\left( \frac{\log(-\square)}{-\Box}  \right)C^{\mu\nu\alpha\beta}
		\end{aligned}
	\end{equation}
	where the explicit form of the gauge-dependent coefficients $\mathcal{C}_{R^2}^\text{Non-loc.}$ and $\mathcal{C}_{C^2}^\text{Non-loc.}$ are given by
	\begin{equation}
		\begin{aligned}
			\mathcal{C}_{R^2}^{Z_N} &= \,
			6 \,g_1 \big[2160 (g_2+1)^2 \lambda ^2 (116 g_2 (g_2+2) (g_3-g_4+1)+116
			g_3-116 g_4+127) \\
			&+7680 (g_2+1)^2 \lambda  \xi  (8 g_2 (g_2+2) (g_3-g_4+1)+8 g_3-8
			g_4+9) \\
			&-\xi ^2 (4 g_2 (g_2 (40 g_2 (38 g_2+125) (g_3-g_4+1) 
			+7 (885 g_3-885 g_4+913 )) \\
			& +3470 g_3-3470 g_4+3796)+5 (604 g_3-604 g_4+717))\big] \\
			&+(g_3-g_4+1)\big[3945600 (g_2+1)^4 \lambda ^3+240 (104 g_2 (10 g_2+17)+1025) (g_2+1)^2 \lambda ^2 \xi \\
			&+40 (8
			g_2 (218 g_2+127)+55) (g_2+1)^2 \lambda  \xi ^2-(16 g_2 (g_2 (8 g_2 (85
			g_2+217)+1671)+773)+2663) \xi ^3\big] \,,
		\end{aligned}
	\end{equation}
	and
	\begin{equation}
		\begin{aligned}
			\mathcal{C}_{C^2}^{Z_N} &= 
			(-g_3+g_4-1) \big[-51840 (g_2+1)^3 (127g_2+139) \lambda ^3-720 (g_2+1)^2 (16 (g_2-34) g_2-227) \lambda ^2 \xi \\
			&+24 (g_2+1) (g_2 (8g_2 (70 g_2+243)+1893)+617) \lambda  \xi ^2+(4 g_2+1)^2 (64 g_2 (g_2+2)+73) \xi^3\big] \\
			&-54g_1 \big[-48 (g_2+1)^2 \lambda ^2 (1276 g_2 (g_2+2) (g_3-g_4+1)+1276 g_3-1276g_4+1461) \\
			&+128 (g_2+1)^2 \lambda  \xi  (8 g_2 (g_2+2) (g_3-g_4+1)+8 g_3-8g_4+9)+\xi ^2 (4 g_2 (g_2 (24 g_2 (2 g_2+7) (g_3-g_4+1)  \\
			&+237 g_3 -237g_4+241)+6 (27 g_3-27 g_4+28))+180 g_3-180 g_4+191)\big] \,.
		\end{aligned}
	\end{equation}
	Particularizing the above result to the minimal gauge, we find
	\begin{equation}
		\begin{aligned}
			\big[\UVB-\IRB\big]_{Z_N}^\text{min.~gauge} 
			&= \frac{ Z_N \left(41868 \lambda ^3+5802 \lambda ^2 \xi -178 \lambda  \xi ^2-21 \xi^3\right)}{10368 \pi ^2 \xi ^2} \, R\left( \frac{\log(-\square)}{-\Box}  \right) R \\
			&-\frac{ Z_N \left(1008 \lambda ^3-4794 \lambda ^2 \xi +73 \lambda  \xi ^2+12 \xi ^3\right)}{20736 \pi ^2 \lambda  \xi} \, C^{\mu\nu\rho\sigma} 
			\left( \frac{\log(-\square)}{-\Box}  \right) C_{\mu\nu\rho\sigma} \,.
		\end{aligned}
	\end{equation}
    We checked that there are no gauge-invariant linear combinations of the momentum-induced beta functions of local and non-local operators, so it is impossible to build any momentum-induced gauge-invariant beta functions associated to the masses $m_2^2$ and $m_0^2$.
	
	\subsection{conformal gravity}
	In conformal gravity, the $\mu$-running comes from
	\begin{equation}
		\UVB + \Tad+ \UVBgh + \UVBY  =-\frac{199}{480 \pi ^2} C^2 \, ,\label{muCG}
	\end{equation}
	which is gauge independent, in agreement with Refs.~\cite{Fradkin:1981iu,deBerredo-Peixoto:2003jda,Ohta:2015zwa}. Thus, the beta function capturing the $\mu$-running of $\lambda$ is given by
    \begin{equation}
        \beta_\lambda^{(\mu)} = -\frac{1}{(4 \pi)^2} \frac{199 \lambda^2}{15} \,.
    \end{equation}
	
	The term proportional to $\log |p|$ is gauge dependent, and the explicit expression is given by
	\begin{equation}
		\begin{aligned}
			&-\Big[ \UVB+\IRB+\UVBgh+\UVBY \Big]=\\ 
			&= \frac{-10 g_1^2 (9 g_3-9 g_4+11)+45 g_1 \lambda  (51 g_3-51 g_4+59)
				(g_3-g_4+1)+19944 \lambda ^2 (g_3-g_4+1)^2}{77760 \pi ^2 \lambda ^2 (g_3-g_4+1)^2} \, C_{\mu\nu\alpha\beta} \log(-\square) C^{\mu\nu\alpha\beta} \, .
		\end{aligned}
	\end{equation}
    Then, the beta function defined according with the $p$-running is
    \begin{equation}
        \beta_\lambda^{(p)} = 
        \frac{-10 g_1^2 (9 g_3-9 g_4+11)+45 g_1 \lambda  (51 g_3-51 g_4+59)
				(g_3-g_4+1)+19944 \lambda ^2 (g_3-g_4+1)^2}{19440 \pi ^2 (g_3-g_4+1)^2} \,,
    \end{equation}
    which turns out to be gauge-dependent.
    
	The minimal gauge in conformal gravity correspond to the following choice of gauge parameters\footnote{This gauge choice can be obtained by talking the limit $\xi \to \infty$ in the minimal gauge of quadratic gravity, \textit{i.e.}, by taking the limit $\xi\to\infty$ in Eq. \eqref{eq:MinGFparameters}.}
	\begin{equation}\label{eq:MinCGGFparameters}
		\begin{split}
			g_1=\lambda\,, \qquad g_2=-\frac14\,, \qquad g_3=\frac23\,, \qquad g_4=1
			\,.
		\end{split}
	\end{equation}
	The $\log |p|$ term reduces to
	\begin{equation}
		-\Big[ \UVB+\IRB+\UVBgh+\UVBY \Big]^\text{min.~gauge} = \frac{93}{320 \pi ^2} \,C_{\mu\nu\alpha\beta} \log(-\square) C^{\mu\nu\alpha\beta}\, .
	\end{equation}
        Thus, the corresponding beta function associated with the $p$-running of $\lambda$ is
        \begin{equation}
            \beta_\lambda^{(p)}\big|_\text{min.~gauge} = -\frac{1}{(4\pi)^2} \frac{93 \lambda^2}{5} \,,
        \end{equation}
	which agrees with the result given in Ref.~\cite{Buccio:2024omv}.

    \section{Analysis}\label{sect:comment}
	   
    The gauge dependence observed in the last section disappears if we go on-shell with the background metric configuration.
	Let us consider the variation of the effective action with respect to the gauge fixing and the Lautrup-Nakanishi operators \cite{Avramidi:1986mj}:
	\begin{equation}\label{eq:varGamma}
		\begin{aligned}
			\delta\Gamma_\text{one-loop}&=-\tr\bigg[-(H^{-1})^{\mu\nu\rho\sigma} \, \epsilon_{\gamma\delta} \, \textbf{R}^{\gamma\delta}{}_{\alpha,\rho\sigma} \,
            (\Delta_{\rm gh}^{-1})^{\alpha\beta} \, \delta F_{\beta,\mu\nu} \\ 
            &+ \frac{1}{2} (\Delta_{\rm gh}^{-1})^{\alpha\beta} \, \epsilon_{\mu\nu} \,\textbf{R}^{\mu\nu}{}_{\alpha,\rho\sigma} \, \left(\textbf{R}^{\rho\sigma}{}_\gamma - (H^{-1})^{\rho\sigma\lambda\delta} \, \epsilon_{\eta\xi} \, \textbf{R}^{\eta\xi}{}_{\gamma,\lambda\delta}\right) (\Delta_{\rm gh}^{-1})^{\gamma\zeta} \, [\delta(Y^{-1})]_{\zeta\beta}
			\bigg]
			\,,
		\end{aligned}
	\end{equation}
    where $\epsilon_{\mu\nu}$ is quadratic gravity equivalent of the Einstein tensor, such that $\epsilon_{\mu\nu} = 0$ corresponds to the vacuum equations of motion in quadratic gravity, $\textbf{R}^{\mu\nu{}}{}_\alpha$ is related to the variations of the metric with respect to a gauge parameter $\xi^\alpha$, namely
        \begin{equation}
		\delta_\xi g^{\mu\nu} = \textbf{R}^{\mu\nu}{}_\alpha\xi^\alpha=\nabla^{(\nu}g^{\mu)}{}_\alpha\xi^\alpha\ ,
	\end{equation}
	$\textbf{R}^{\mu\nu}{}_{\alpha,\rho\sigma}$ is its functional derivative with respect to the metric, and $F_{\alpha,\mu\nu}$ is the functional derivative of the gauge fixing $F_\alpha$ with respect to the fluctuation field $h_{\mu\nu}$.
	The equations of motion have mass dimension 4, because the metric is taken dimensionless.
	Since the expression in the first line of (\ref{eq:varGamma}) can be reduced at the leading order to the inverse of a $4\text{th}$ order differential operator contracted with $\epsilon_{\gamma\delta}$, its UV divergence is given by a tadpole diagram and must be proportional to the trace of $\epsilon$ \cite{Barvinsky:1985an,Avramidi:1986mj}. In the second line, the first term has the same type of divergence, while the second term is composed of two equations of motion multiplied by the inverse of a differential operator of rank $8$, so it is not UV divergent at all.
	That means the gauge dependence of UV divergences is proportional to the trace of the equations of motion.
	\begin{equation}
	   \begin{aligned}
	       \epsilon^{\mu\nu} &= Z_N(R^{\mu\nu}-\frac12 R g^{\mu\nu}) - \frac{1}{\xi}\left(2 R^{\mu \nu} R -  \frac{1}{2} g^{\mu \nu} R^2 - 2 \nabla^{\nu}\nabla^{\mu}R + 2
		g^{\mu \nu} \nabla_{\rho}\nabla^{\rho}R\right) \\
		&+ \frac{1}{\lambda} \left(2 R^{\mu \rho} R^{\nu}{}_{\rho} -  \frac{1}{2}g^{\mu \nu} R_{\rho \sigma} R^{\rho \sigma} -  \frac{1}{3} R^{\mu \nu} R + \frac{1}{12} g^{\mu \nu}R^2 -  R^{\mu \rho \sigma \lambda} R^{\nu}{}_{\rho \sigma \lambda} \right. \\
        &\left.\qquad + \frac{1}{4} g^{\mu \nu} R_{\rho \sigma \lambda \delta} R^{\rho \sigma \lambda \delta} + \frac{1}{3} \nabla^{\nu}\nabla^{\mu}R
	-  \nabla_{\rho}\nabla^{\rho}R^{\mu \nu} + \frac{1}{6} g^{\mu \nu}
		\nabla_{\rho}\nabla^{\rho}R \right) \,.
	   \end{aligned}
	\end{equation}
	The trace of this expression is
	\begin{equation}
		\epsilon=  -Z_N R-\frac{6}{\xi} \, \square R\ ,
	\end{equation}
	therefore only the $\mu$-running of the EH term and the boundary term $\square R$, neglected in our computation, can be gauge dependent \cite{Avramidi:1986mj,Voronov:1984kq}.
	
	When we look for IR divergences, various contractions of $\epsilon_{\mu\nu}$ with other curvature invariants and $\epsilon^{\mu\nu}\epsilon_{\mu\nu}$ can appear, hence the momentum running of the local curvature square terms is not gauge invariant, as explicitly shown above.
	
	To extract from the effective action a set of gauge-independent physical observables, we have to go on-shell.
	
	We have done our computation in the high energy, almost flat regime, that means $\nabla\nabla \mathcal{R}\gg \mathcal{R}^2\gg Z_N \mathcal{R}$. Thanks to this assumption, we have neglected in the effective action higher curvature nonlocal terms $O(\mathcal{R}^3)$.
	In this regime, the equations of motion reduce to
	\begin{subequations}
	    \begin{align}
		&\square R=0\\
		&\frac{2}{\xi}\left(\nabla^{\nu}\nabla^{\mu}R+g^{\mu \nu} \nabla_{\rho}\nabla^{\rho}R\right)+\frac{1}{\lambda}\left(\frac13\nabla^{\nu}\nabla^{\mu}R-  \nabla_{\rho}\nabla^{\rho}R^{\mu \nu} + \frac{1}{6} g^{\mu \nu}
		\nabla_{\rho}\nabla^{\rho}R\right)=O(\mathcal{R}^2) \label{eq:eom2}
	\end{align}
	\end{subequations}
	in the trace and non-trace parts.
	Neither of these equations can directly affect the one-loop terms.
	However, we can make the second equation nonlocal by acting on both sides with $\square^{-1}$, and then contract it with $R_{\mu\nu}$, finding
	\begin{equation}
		\frac{1}{\lambda}R_{\mu\nu}R^{\mu\nu}=\frac{2}{\xi}R^2+\frac{1}{3\lambda}R^2+O\left(\mathcal{R}\frac1\square \mathcal{R}^2\right)\ .
	\end{equation}
	In the Weyl basis, it is equivalent to
	\begin{equation}
		\frac{1}{2\lambda}C^2=\frac{2}{\xi}R^2+O\left(\mathcal{R}\frac1\square \mathcal{R}^2\right)\ .\label{eq:CR}
	\end{equation}
	
	We observe that the scalar equation $\square R=0$, together with asymptotically flat boundary conditions, implies $R=0$ in the absence of external sources, since oscillatory solutions are not 0 at infinity and exponentially growing ones are singular at the origin. This implies $C^2= O\left(\mathcal{R}\frac1\square \mathcal{R}^2\right) $ through (\ref{eq:CR}).
    We can also replicate the same argument by acting on \eqref{eq:eom2} with $\log(-\square)/(-\square)$ and then contracting with $R^{\mu\nu}$ to show that $ C_{\mu\nu\alpha\beta} \log(-\square) C^{\mu\nu\alpha\beta}$ vanishes on-shell within our approximation.
    Hence, the whole one-loop effective action is equal to zero, up to $O\left(\mathcal{R}\frac1\square \mathcal{R}^2\right)$ terms.
	
	In conformal gravity, we keep only the term proportional to $1/\lambda$ in the last equation, so the same result holds even in this case.

	In conclusion, if we go on-shell with the background, we can rewrite all operators quadratic in curvatures in terms of operators cubic in curvatures, which we neglected in our computation, so the effective action becomes zero at our order of approximation.
	This is a manifestation of the fact that the field renormalization is, in general, unphysical and gauge-dependent, since it does not correspond to an on-shell scattering amplitude.
	
	In Yang-Mills theories and in the UV divergent part of quadratic gravity and conformal gravity, the background field method \cite{Weinberg:1996kr} permits one to extract from it the running of the couplings governing, for example, the 4-point amplitude, the simplest on-shell non-trivial process. This happens because the resulting theory for the background field is invariant with respect to background gauge transformations, and the UV divergences have to be strictly local.
	These two constraints, together with dimensional analysis, fix the structure of the YM background one-loop action to only $F^2$, the UV divergent part of the quadratic gravity action to the form $R^2+C^2$, and that of conformal gravity to only $C^2$. The UV divergent parts of both the 2-point function and the 4-point amplitude come from these operators, hence, they must be governed by the same coupling.
	
	If we consider also IR effects, absent in YM, since it is a 2-derivative theory, any nonlocal covariant operator of dimension 4 can be generated in the effective action. These operators have the general structure
	\begin{equation}
		\mathcal{R}\square^{-n}\mathcal{R}^{n+1}\ .\end{equation}
	If $n=1,2$, they contribute at one-loop to the 4-point on-shell scattering, but not to the background propagator. In this case, the resulting on-shell scattering amplitude is not proportional to the vertex generated by the local terms $R^2+C^2$ (or only $C^2$ in conformal gravity), and its dependence on momenta is not completely determined by their beta functions, even when taking into account IR contributions. The beta functions of the local operators, determined by the momentum dependence of the 2-point functions, are no longer associated with an on-shell gauge-invariant process, and can show a gauge dependence, as explicitly observed above and in \cite{Salvio:2025cmi}. That means the beta functions of the local operators are not enough to establish the UV behaviour of the theory, contrary to what happens in YM theories.
	To have something gauge invariant, we must compute on-shell one-loop scattering amplitudes, as done in \cite{Salvio:2025cmi} for the scalar sector of quadratic gravity.

    \section{Conclusions}\label{sect:conclusions}
	In accordance with \cite{Salvio:2025cmi}, we have found that the momentum-induced beta functions, which some of us claimed to be physical in previous works \cite{Buccio:2024hys,Buccio:2024omv}, carry explicit dependence on unphysical parameters such as the $g_i$'s. This gauge dependence is due to the presence of logarithmically enhanced nonlocal operators in the one-loop effective action. They contribute to the on-shell $n$-point scattering amplitudes with $n>2$, which are the truly physical objects, and mix with local ones. Only the resulting combination of contributions needs to be gauge independent, not the single beta functions taken one by one.
	
	In light of these results, the explicit computation of on-shell scattering amplitudes in quadratic gravity is necessary to understand the UV regime of quadratic gravity and whether it is an asymptotically free theory. Both versions of the beta functions of the local operators $R^2$ and $C^2$, defining respectively the $\mu$-running and the momentum running, are not sufficient to determine the leading UV behaviour of all the scattering amplitudes of the theory.
	
	In \cite{Salvio:2025cmi}, the 4-point amplitudes for the scalar mode were computed and were found to be also parametrization dependent. This, together with the presence of infrared divergences, could be the signal that the canonical degrees of freedom of quadratic gravity are not the correct asymptotic states of the theory in the high-energy regime. It could be that considering inclusive external states where ghost and massless particles are mixed, as suggested in \cite{Holdom:2021hlo}, cancels the IR large logs from scattering amplitudes.

	\begin{acknowledgments}
		
	We thank A. Salvio, A. Strumia, J. F. Donoghue, G. Menezes, R. Percacci, O. Zanusso, M. Reichert, L. Buoninfante, W. Cesar e Silva, I. Shapiro and A. Lehum for useful discussions. We also thank R. Percacci and O. Zanusso for comments on the manuscript.
	D.B. and L.P. thank Frank Saueressig and IMAPP for the hospitality. D.B. acknowledges Fondazione Angelo Della Riccia for the financial support.
		
	\end{acknowledgments}

	\bibliographystyle{apsrev4-2}
	\bibliography{biblio}
	
\end{document}